\documentclass[twocolumn,english,prb,epsfig,rotate,showpacs,aps]{revtex4-1}
\usepackage[T1]{fontenc}
\usepackage[latin9]{inputenc}
\usepackage{color}
\usepackage{verbatim}
\usepackage{amsmath}
\usepackage{amssymb}
\usepackage{graphicx}
\usepackage{esint}
\usepackage{microtype}
\usepackage{hyperref}
\usepackage{lmodern}
\makeatletter
\usepackage{babel}

\usepackage{epsfig}
\usepackage{amsmath}
\usepackage{graphicx}
\usepackage{dcolumn}
\usepackage{bm}
\usepackage{amsfonts,amssymb}
\usepackage[T1]{fontenc}
\usepackage[latin9]{inputenc}
\usepackage{amsmath}
\usepackage{graphicx}
\usepackage{amssymb}
\usepackage{esint}
\usepackage{babel}
\usepackage{microtype}
\usepackage{xcolor}
\usepackage{ulem}

\makeatother

\usepackage{babel}

\begin{document}

\setlength{\baselineskip}{11pt} 
\title{Intrinsic superconducting diode effects in tilted Weyl and Dirac semimetals}
\author{Kai Chen }
\affiliation{Department of Physics and Texas Center for Superconductivity, University
of Houston, Houston, TX 77204}
\author{Bishnu Karki}
\affiliation{Department of Physics and Texas Center for Superconductivity, University
of Houston, Houston, TX 77204}
\author{Pavan Hosur}
\affiliation{Department of Physics and Texas Center for Superconductivity, University
of Houston, Houston, TX 77204}
\date{\today}
\begin{abstract}
We explore Weyl and Dirac semimetals with tilted nodes as platforms
for realizing an intrinsic superconducting diode effect. Although
tilting breaks sufficient spatial and time-reversal symmetries, we
prove that -- at least for conventional $s$-wave singlet pairing
-- the effect is forbidden by an emergent particle-hole symmetry
at low energies if the Fermi level is tuned to the nodes. Then, as
a stepping stone to the three-dimensional semimetals, we analyze a
minimal one-dimensional model with a tilted helical node using Ginzburg-Landau
theory. While one might naively expect a drastic enhancement of the
effect when the node turns from type-I to type-II, we find that the
presence of multiple Fermi pockets is more important as it enables
multiple pairing amplitudes with indepedent contributions to supercurrents
in opposite directions. Equipped with this insight, we construct minimal
lattice models of Weyl and Dirac semimetals and study the superconducting
diode effect in them. Once again, we see a substantial enhancement
when the normal state has multiple Fermi pockets per node that can
accommodate more than one pairing channel. In summary, this study
sheds light on the key factors governing the intrinsic superconducting
diode effect in systems with asymmetric band structures and paves
the way for realizing it in topological semimetals.
\end{abstract}
\maketitle

\section{Introduction}

In recent years, there has been a growing interest in the field of
electronics and superconductivity due to the fascinating observation
of superconducting diode effects (SDEs). These effects involve the
ability of certain materials and structures to exhibit nonreciprocal
superconducting transport, effectively blocking electric current flow in
one direction while allowing it to pass in the opposite direction.
This behavior resembles that of a diode, making SDEs crucial for devising
rectifiers and switches.

A seminal experimental study by Ando et al. \citep{ando2020observation}
demonstrated the presence of SDEs in an artificial superlattice {[}Nb/V/Ta{]}.
This observation was achieved by breaking the inversion symmetry of
the structure and introducing time-reversal symmetry breaking through
the application of an external magnetic field. Since then, the study
of SDEs has become an active area of research in the field of superconductivity,
owing to the significant potential of nonreciprocal critical supercurrent
in various applications, such as electronics, spintronics, phase-coherent
charge transport, direction-selective charge transport, and quantum
computation using superconductor qubits \citep{linder2015superconducting,golod2022demonstration,pal2022josephson,jiang2022superconducting,nadeem2023superconducting,narita2022field}.

Experimental investigations have explored SDEs in diverse materials
and structures. For instance, SDEs have been observed in magic angle
twisted graphenes \citep{lin2022zero,scammell2022theory,diez2023symmetry}, in few layer $\text{NbSe}_2$ \citep{bauriedl2022supercurrent}.
Furthermore, Josephson supercurrent diode effects have been demonstrated in highly transparent Josephson junctions fabricated on InAs quantum wells \citep{baumgartner2022supercurrent},
in van der Waals heterostructures and symmetric Al/InAs-2DEG/Al junctions \citep{wu2022field},
in a three-terminal Josephson device based upon an InAs quantum well \citep{gupta2023gate}
and Josephson junctions containing single magnetic atoms \citep{trahms2023diode}. The thin superconducting films made of niobium and vanadium indicate
a robust SDE when exposed to an extremely low magnetic field of 1
Oe. Furthermore, when a layer of EuS is introduced, the SDE is amplified \citep{hou2023ubiquitous}. For asymmetric vortex motion, which exposes the mechanism underpinning the superconducting vortex
diode phenomenon, has been reported in the layered structure of Nb/EuS
(superconductor/ferromagnet) \citep{gutfreund2023direct}. SDE has also been observed in topological insulator/superconductor \citep{yasuda2019nonreciprocal, masuko2022nonreciprocal, karabassov2022hybrid} and superconductor nanowire/topological Dirac semimetal \citep{ishihara2023giant} hybrid systems.

The intriguing experimental findings have stimulated theoretical efforts
to understand the underlying mechanisms of SDEs. The Rashba-Zeeman-Hubbard
model has been proposed as a theoretical framework to explain SDEs,
and established a close relationship between SDE and Fulde-Ferrell-Larkin-Ovchinnikov (FFLO) states \citep{yuan2022supercurrent,daido2022intrinsic}. In the FFLO state, Cooper pairs form with finite center-of-mass momenta due to opposite spin states on Zeeman-split Fermi surfaces \citep{fulde1964superconductivity,larkin1965nonuniform}.
Numerical calculations and Ginzburg-Landau (GL) theory have provided
further support and insights into the understanding of SDEs \citep{daido2022intrinsic,daido2022superconducting}. Among extrinsic mechanisms, SDE behavior has been predicted in topological insulators and Rashba nanowires \citep{legg2022superconducting} as well as general metallic wires with asymmetric dispersion, with the latter expected to show the theoretically maximum SDE in a range of parameters \citep{hosur2022equilibrium}. Moreover, researchers have investigated the influence of disorder on SDEs by using the quasi-classical Eilenberger equation \citep{ilic2022theory}.
The disorder effect is crucial in comprehending the behavior of SDEs
in realistic and practical scenarios. Theoretical studies have also
focused on the Josephson diode effect, revealing its universality
and potential applicability in various contexts \citep{zhang2022general,davydova2022universal,zhang2022general,souto2022josephson,legg2022superconducting,wang2022symmetry}.

This work explores intrinsic SDEs in Weyl and Dirac semimetals. These semimetals are characterized by gapless points between their valence and conduction bands, known as Weyl and Dirac points, respectively \citep{young2012dirac,young2015dirac,gibson2015three,hosur2013recent,yan2017topological,armitage2018weyl}. They possess several favorable properties that make them promising platforms for the SDEs. For instance, the density of states near the nodes is low, which facilitates breaking of time-reversal, inversion and spatial symmetries necessary for enabling the SDE. These materials also typically have multiple Fermi pockets centered at different points in momentum space, which enhances the possibility of FFLO states \citep{cho2012superconductivity,wei2014odd,bednik2015superconductivity,Hao2017}. Moreover, Fermi pockets centered around the origin can also develop finite momentum pairing if the dispersion is tilted. There are two different types of Weyl/Dirac semimetals: type I, with point-like Fermi surfaces, and type II, defined by electron and hole pockets touching at the Weyl nodes \citep{zyuzin2016intrinsic,tchoumakov2016magnetic,soluyanov2015type}. Tilting the dispersion around the node induces the transition from type-I to type-II. In this study, we shed light on the key factors that enhance the SDE in tilted semimetals. In particular, we show that multiple inequivalent pairing channels can enhance the intrinsic SDEs and are more important than the band tilting.

The outline of this paper is as follows. In Section II, we delve into the symmetries beyond time reversal and inversion symmetry that need to be broken in order to support SDEs. We explore
how tuning the chemical potential impacts these symmetries, shedding
light on the underlying symmetry breaking responsible for SDEs and
offering potential avenues for experimental control and manipulation
of these effects. In Section III, we employ the Ginzburg-Landau theory to investigate
a one-dimensional model characterized by an asymmetric band structure.
Our analysis reveals that this simple yet insightful model can indeed
support a ground state with Cooper pairs possessing finite momentum,
thus providing a compelling platform to observe and study SDEs. Building on the insights gained from the 1D model, we extend our study to lattice modes of tilted Weyl semimetals and Dirac semimetals in sections IV and V, respectively. Our numerical simulations reveal the existence of nonreciprocity in the depairing critical current, the key requirement for SDEs in these intriguing materials, and support the heuristic that multiple inequivalent pairing channels are more important than band asymmetry for a large SDE.

\section{Symmetry and the role of chemical potential $\mu$\label{sec:symmetry}}

In general, necessary conditions for realizing the SDE are the violation
of time-reversal ($\mathcal{T}$), inversion ($\mathcal{I}$) and
spatial symmetries under which current in the desired non-reciprocal
direction is odd. These conditions ensure the breaking of reciprocity
in the system, meaning that the response of the superconductor to
external perturbations is different for perturbations applied in opposite
directions. In most cases, these violations suffice to guarantee a
SDE; however, a chiral or particle hole symmetry in the normal state,
common found at low energies near band intersections, can suppress
the SDE for singlet pairing as shown below. 

Consider a Bloch Hamiltonian $H(\mathbf{k})$. The Bogoliubov-de Gennes
(BdG) Hamiltonian for generic pairing in the basis $\left(c_{\mathbf{k}+{\mathbf{q}/2}},c_{-\mathbf{k}+\mathbf{q}/2}^{\dagger}\right)^{T}$
is 
\begin{equation}
H^{\text{BdG}}(\mathbf{k},\mathbf{q},\Delta_{\mathbf{k}})=\begin{pmatrix}H(\mathbf{k}+\mathbf{q}/2) & \Delta_{\mathbf{k}}\\
\Delta_{\mathbf{k}}^{\dagger} & -H^{*}(-\mathbf{k}+\mathbf{q}/2)
\end{pmatrix}
\end{equation}
where we have allowed for pairing with finite momentum $\mathbf{q}$
and fermion antisymmetry ensures $\Delta_{\mathbf{k}}=-\Delta_{-\mathbf{k}}^{T}$.
$H^{\text{BdG}}(\mathbf{k},\mathbf{q},\Delta)$ obeys particle-hole
symmetry 
\begin{equation}
\tau_{x}\mathbb{K}H^{\text{BdG}}(\mathbf{k},\mathbf{q},\Delta_{\mathbf{k}})\mathbb{K}\tau_{x}=-H^{\text{BdG}}(-\mathbf{k},\mathbf{q},\Delta_{\mathbf{k}})
\label{eq:particle-hole}
\end{equation}
where $\tau_{x}$ is a Pauli matrix in Nambu space and $\mathbb{K}$
denotes complex conjugation.

Suppose the normal state also has a chiral unitary symmetry $Q$:
\begin{equation}
QH(\mathbf{k})Q^{\dagger}=-H(\mathbf{k})\label{eq:chiral-unitary}
\end{equation}
or a chiral anti-unitary or particle-hole symmetry $Q\mathbb{K}$:
\begin{equation}
Q\mathbb{K}H(\mathbf{k})\mathbb{K}Q^{\dagger}=-H^{*}(-\mathbf{k})\label{eq:chiral-anti}
\end{equation}
Under $Q$ and $Q\mathbb{K}$, $H^{\text{BdG}}(\mathbf{k},\mathbf{q},\Delta_{\mathbf{k}})$
transforms into $-H^{\text{BdG}}(\mathbf{k},\mathbf{q},-\tilde{\Delta}_{\mathbf{k}})$
and $-\tau_{x}H^{\text{BdG}}(-\mathbf{k},\mathbf{q},\tilde{\Delta}_{\mathbf{k}})\tau_{x}$,
respectively, where $\tilde{\Delta}_{\mathbf{k}}=Q\Delta_{\mathbf{k}}Q^{\dagger}$.
Along with the BdG particle-hole symmetry Eq. (\ref{eq:particle-hole}),
these two symmetries in the normal state ensure that $H^{\text{BdG}}(\mathbf{k},\mathbf{q},\Delta_{\mathbf{k}})$
is related to $H^{\text{BdG}}(\mathbf{k},-\mathbf{q},-\tilde{\Delta}_{\mathbf{k}})$
and $H^{\text{BdG}}(-\mathbf{k},-\mathbf{q},\tilde{\Delta}_{\mathbf{k}})$
by anti-unitary and unitary operations.

Assuming the electrons experience an attractive Hubbard interaction
($g>0$) 
\begin{equation}
H_{\text{int}}=-g\sum_{\mathbf{k},\mathbf{k}',\mathbf{q}}c_{\mathbf{k}+\frac{\mathbf{q}}{2}\uparrow}^{\dagger}c_{-\mathbf{k}+\frac{\mathbf{q}}{2}\downarrow}^{\dagger}c_{-\mathbf{k}'+\frac{\mathbf{q}}{2}\downarrow}c_{\mathbf{k}'+\frac{\mathbf{q}}{2}\uparrow},
\end{equation}
where $g$ represents 
the strength of attraction. Within the mean field approximation, we
get the Ginzburg-Landau free energy density: 
\begin{equation}
f[\mathbf{q},\Delta]=\intop_{\mathbf{k}}\frac{\text{tr}(\Delta_{\mathbf{k}}\Delta_{\mathbf{k}}^{\dagger})}{g}-T\text{Tr}\log\left[1+e^{-H_{\text{BdG}}(\mathbf{k},\mathbf{q},\Delta_{\mathbf{k}})/T}\right]
\end{equation}
where $\intop_{\mathbf{k}}\equiv\int\frac{d^Dk}{(2\pi)^D}$, $D$ is the spatial dimension of the system, $\text{tr}\left(\dots\right)$ runs over spin and orbitals while
$\text{Tr}\left[\dots\right]$ runs over spin, orbital and Nambu degrees
of freedom. Clearly, $f(\mathbf{q},\Delta)$ only depends on the energy
eigenvalues of $H_{\text{BdG}}(\mathbf{k},\mathbf{q})$ and is unaffected
under the change $\mathbf{k}\to-\mathbf{k}$ of the integration variable.
Moreover, $U(1)$ gauge symmetry mandates $f(\mathbf{q},\Delta)$
to be unchanged under the transformation $\Delta_{\mathbf{k}}\to e^{i\phi_{\mathbf{k}}}\Delta_{\mathbf{k}}$
for arbitrary $\phi_{\mathbf{k}}$. Thus, if $\Delta_{\mathbf{k}}$
equals $\tilde{\Delta}_{\mathbf{k}}$ ($\tilde{\Delta}_{-\mathbf{k}}$)
upto a phase when the normal state possesses the symmetry $Q$ ($Q\mathbb{K}$),
$f(\mathbf{q},\Delta)$ is even in $\mathbf{q}$: $f(\mathbf{q},\Delta)=f(-\mathbf{q},\Delta)$.
The above condition on $\Delta_{\mathbf{k}}$ is clearly obeyed by
ordinary spin singlet $s$-wave pairing, $\Delta_{\mathbf{k}}=\Delta\sigma_{y}$
with $\sigma_{y}$ a spin Pauli matrix. Henceforth, we take pairing
to be of this form and assume $\Delta_{\mathbf{k}}\equiv\Delta$ independent
of $\mathbf{k}$. Note, $\Delta$ can still pair electrons with non-zero
center-of-mass momentum $\mathbf{q}/2$.

The SDE can be calculated by minimizing $f[\mathbf{q},\Delta]$ with
respect to $\Delta$ for fixed $\mathbf{q}$ to obtain the condensation
energy $f[\mathbf{q},\Delta(\mathbf{q})]\equiv f(\mathbf{q})$ at
that $\mathbf{q}$, followed by extremizing the supercurrent $j(\mathbf{q})\equiv2\partial_{\mathbf{q}}f(\mathbf{q})$
over $\mathbf{q}$. Positive and negative currents of largest magnitudes
represent critical currents in opposite directions, $j_{c}^{\pm}$,
and the SDE is characterized by the quality factor
\begin{equation}
\eta=\left|\frac{j_{c}^{+}-j_{c}^{-}}{j_{c}^{+}+j_{c}^{-}}\right|\in[0,1]
\end{equation}
If $f(\mathbf{q})=f(-\mathbf{q})$, critical currents in opposite
directions have the same magnitude and the SDE is absent ($\eta=0$)
while the largest SDE occurs if either $j_{c}^{+}$ or $j_{c}^{-}$
vanishes.

Point nodes in band structures enjoy at least one of chiral or particle-hole
symmetries at low energies when the chemical potential is tuned to
the node. For instance, in the absence of tilting, massless 2D Dirac
nodes enjoy the chiral symmetry $Q$, 3D Weyl nodes respect $Q\mathbb{K}$,
and 3D Dirac nodes possess both $Q$ and $Q\mathbb{K}$. Crucially,
while $Q$ is immediately violated by a tilt in the dispersion, $Q\mathbb{K}$
survives. Therefore, to obtain a SDE with $s$-wave, singlet pairing
in tilted Weyl and Dirac semimetals, the chemical potential must be
tuned away from the node to break the particle-hole symmetry $Q\mathbb{K}$
in the normal state.

Note that a finite chemical potential is not merely a density of states
requirement for superconductivity to occur in the first place. Indeed,
type-II semimetals already possess finite Fermi surfaces and hence,
a superconducting instability with appropriate interactions. Instead,
a finite chemical potential is symmetry requirement for the SDE that
goes beyond the usual mandates of broken $\mathcal{T}$, $\mathcal{I}$
and other spatial symmetries that reverse the supercurrent.

\section{SDE in a minimal 1D model with asymmetric bands\label{sec:1D-model}}

In this section, we focus on a one-dimensional (1D) model with asymmetric
bands. This will yield insight that will be useful for understanding
the SDE for 3D Weyl and Dirac fermions. In particular, we will gradually
develop the following intuition: when multiple pairing channels are
present, it is possible for critical currents in opposite directions
to be dominated by different channenls and can therefore be vastly
different, resulting in a large SDE.

\begin{figure}[h]
\includegraphics[width=0.95\columnwidth]{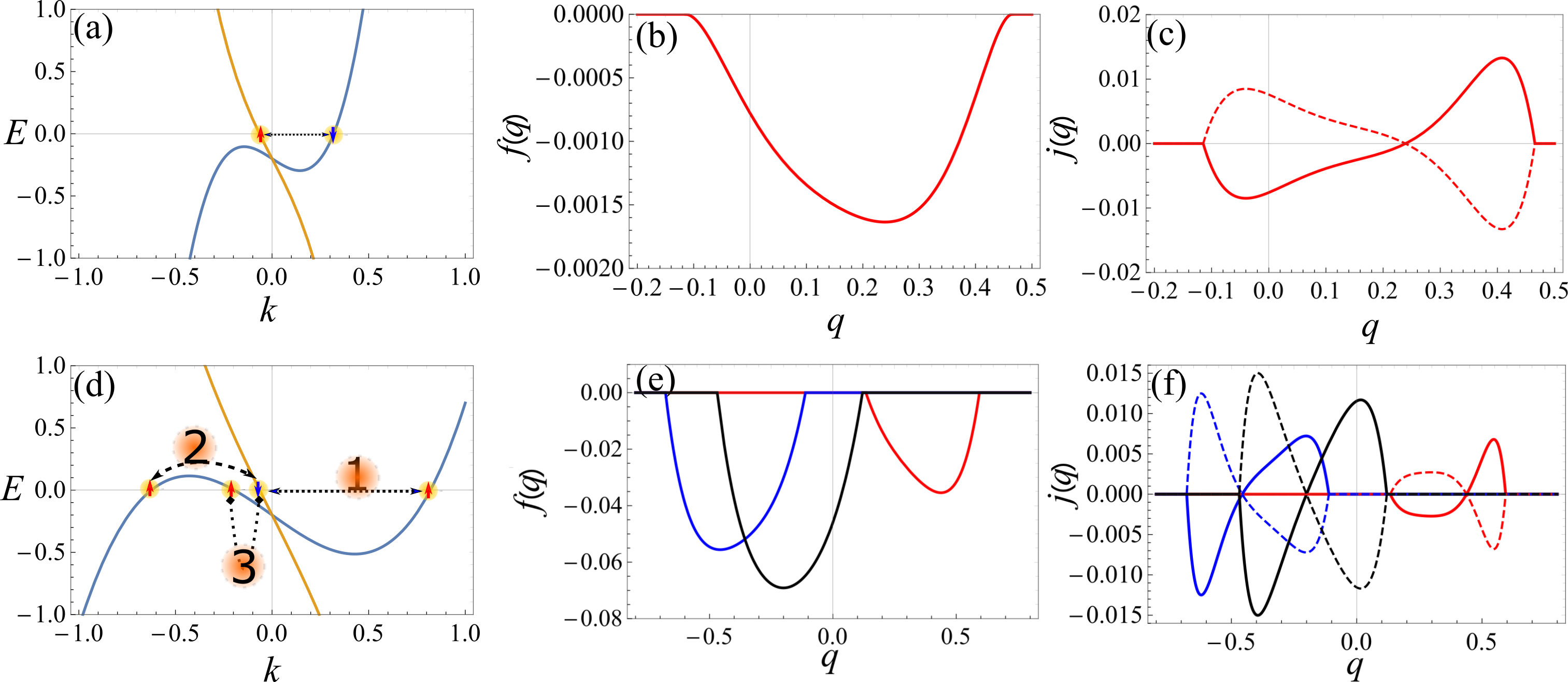} \caption{(a) and (d): Schematics of Cooper pairs in the quasi-one-dimensional
system. (b) and (c): The GL free energy density $f(q)$, the supercurrent
$j(q)$ (solid line), and $-j(q)$ (dashed line) for weak tilting
with $\lambda=2$. The parameters are $\alpha=16$, $\lambda=1$,
$T_{c}\approx0.46$, $T=0.1$, and $\mu=0.4$. (e): The GL free energy
density $f(q)$ for different Cooper pairs: red line (Cooper pairing
channel 1), blue line (Cooper pairing channel 2), and black line (Cooper
pairing channel 3). The supercurrent $j(q)$ for different Cooper
pairs: red line (Cooper pairing channel 1), blue line (Cooper pairing
channel 2), and black line (Cooper pairing channel 3). Dashed lines
represent the opposite supercurrent $-j(q)$ for Cooper pairing channels
with the same color. The parameters in (e) and (f) are the same as
in (b) and (c), except that the parameter $\lambda=4.4$.}
\label{fig:1D-analytic} 
\end{figure}

A minimal model can be described by
\begin{equation}
H_{1D}(k)=(1+\alpha k^{2})k\sigma_{z}-\lambda k-\mu,\label{eq:h1D}
\end{equation}
where $\mu$ is the chemical potential and $\sigma_{z}$ is the Pauli-Z
matrix in spin space. The parameter $\lambda$ creates a tilt in the
dispersion around $k=0$ while $\alpha>0$ ensures that the tilt is
undone at finite $k$. $H_{1D}$ has two qualitatively different regimes
separated by a critical value of $\lambda$, 
\begin{equation}
\lambda_{c}=\left|1+3\left(\frac{\mu^{2}|\alpha|}{4}\right)^{1/3}\right|
\end{equation}
for given $\alpha$ and $\mu$. For $|\lambda|<\lambda_{c}$, there
are only two Fermi points and one momentum channel for Cooper pairing,
while $|\lambda|>\lambda_{c}$ results in four Fermi points and three
channels as sketched in Fig. \ref{fig:1D-analytic}(a,d). 

For singlet superconductivity with Cooper pair momentum $q$, the
appropriate BdG Hamiltonian is 
\begin{equation}
H_{1D}^{\text{BdG}}(k,q)=\begin{pmatrix}H_{1D}(k+q/2) & -i\sigma_{y}\Delta\\
i\sigma_{y}\Delta & -H_{1D}^{*}(-k+q/2).
\end{pmatrix},
\label{eq:1D-HBdG}
\end{equation}
At $\mu=0$, $H_{1D}$ satisfies a particle-hole symmetry, $\sigma_{y}H_{1D}^{*}(k)\sigma_{y}=-H_{1D}(-k)$,
which suppresses the SDE as described in Sec. \ref{sec:symmetry}
with $Q\equiv\sigma_{y}$. At non-zero $\mu$, we calculate the diode
coefficient $\eta$ in three different ways with increasing amount
of analytical input and physical insight.

\begin{figure}[h]
\includegraphics[width=0.95\columnwidth]{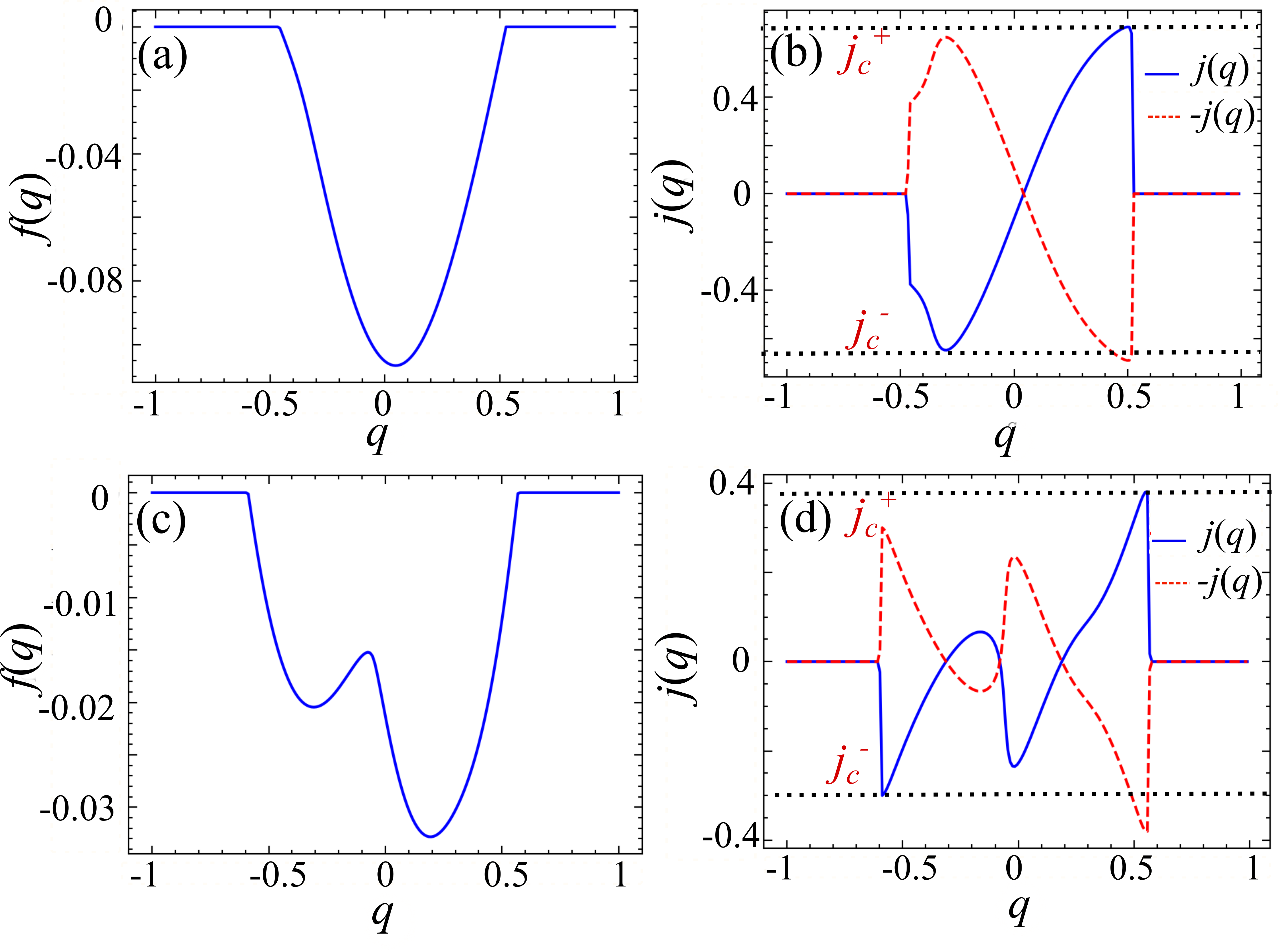} \caption{(a, b): Free energy density and supercurrent with parameter $\lambda=2$.
(c, d): Free energy density and supercurrent with parameter $\lambda=4.4$.
Other parameters are $\alpha=16$, $\mu=0.4$ and $g=3\pi$, which yield
$\lambda_{c}\approx3.58$ and $T_{c}\approx0.46$, and we set $T=0.1$.}
\label{fig:1D-numeric}
\end{figure}

First, we directly compute the free energy density

\begin{equation}
f[q,\Delta]=\frac{|\Delta|^{2}}{g}-T\int\frac{dk}{2\pi}\text{Tr}\log\left(1+e^{-\frac{H_{\text{BdG}}^{1D}(k,q)}{T}}\right),\label{eq:1D-fq}
\end{equation}
minimize it with respect to $\Delta$ to obtain $\Delta(q)$ upto a phase and
$f(q)\equiv f[q,\Delta(q)]$, and compute the current $j(q)=2\partial_{q}f(q)$.
All steps are carried out numerically and the results are shown in
Fig. \ref{fig:1D-numeric}. For weak tilting, $|\lambda|<\lambda_{c}$,
we see a single minimum in $f(q)$ close to $q=0$ and a small diode
coefficient $\eta\approx3.2\%$ {[}Fig. \ref{fig:1D-numeric}(a,b){]}.
Strong tilting unsurprisingly produces a larger $\eta\approx12\%$.
However, the enhancement is not merely quantitative; we observe qualitatively
new features in $f(q)$ in the form of two inequivalent local minima
away from $q=0$ and a large corresponding asymmetry in $j(q)$ {[}Fig.
\ref{fig:1D-numeric}(c,d){]}, suggesting that the change in Fermiology
plays an important role in enhancing the SDE.

\begin{figure}
\includegraphics[width=0.95\columnwidth]{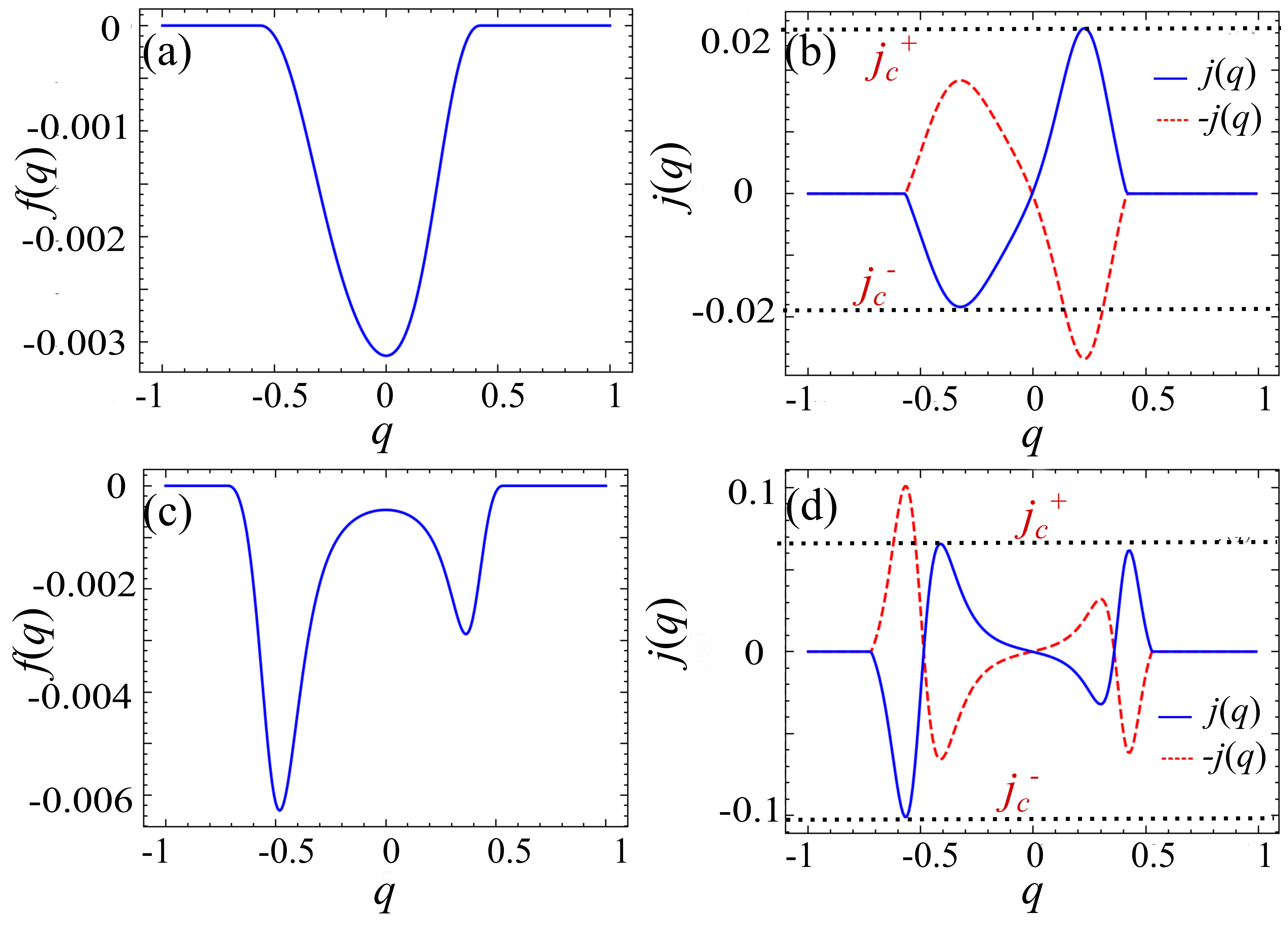}

\caption{(a), (b): The GL free energy density $f(q)$ and the supercurrent
$j(q)$ (blue line) and $-j(q)$ (red line) under weak tilting with
$\lambda=2$, respectively. (c), (d): The same quantities as (a, b)
under strong tilting with $\lambda=4.4$, respectively. The parameters
are $\alpha=16$, $T_{c}\approx 0.46$, $T=0.1$, and $\mu=0.4$.
}\label{fig:1D-Aq-numeric} 
\end{figure}

To analyze this point further, we focus on $T$ close to the critical
temperature $T_{c}$ where $\Delta$ is small and $f[q,\Delta]$ can
be approximated as 
\begin{equation}
f[q,\Delta]=A(q)\Delta^{2}+\frac{B(q)}{2}\Delta^{4},\label{eq:1D-fq-GL}
\end{equation}
In this regime, the main role of $B(q)$ is to ensure physical stability
by lower bounding $f[q,\Delta]$, allowing us to safely take it to
be a positive constant, $B(q)\approx b>0$, (we set $b=1$ throughout this work).
In contrast, the physics of the system depends sensitively on $A(q)$.
For instance, minimizing $f[q,\Delta]$ yields a superconducting ground
state with $\left|\Delta(q)\right|=\sqrt{-A(q)/b}$ only if $A(q)<0$,
while the supercurrent an be expressed as $j(q)=2\frac{\partial}{\partial q}f(q)=|A(q)|\frac{\partial}{\partial q}A(q)$.
Thus, we explicitly calculate $A(q)$ following \citep{he2022phenomenological}
as: 
\begin{align}
A(q) & =-T\int\frac{dk}{2\pi}\sum_{n}\text{tr}[G(k+q,\epsilon_{n})G(-k,-\epsilon_{n})]\label{eq:Aq}\\
 & +T_{c}\int\frac{dk}{2\pi}\sum_{n}\text{tr}[G(k,\epsilon_{n})G(-k,-\epsilon_{n})]_{T=T_{c}},\nonumber 
\end{align}
where the Matsubara Green's function $G(k,\epsilon_{n})=[i\epsilon_{n}-H_{1D}(k)]^{-1}$
with $\epsilon_{n}=(2n+1)\pi T$. The second term in Eq. \ref{eq:Aq} reduces to just $1/g$, which determines the value of the critical temperature $T_c$. The momentum integral is carried
out numerically and $A(q)$ hence obtained is used to reevaluate $f(q)$
using Eq. \ref{eq:1D-fq-GL}. The results, shown
in Fig. \ref{fig:1D-Aq-numeric}, are qualitatively consistent with
the fully numerical results presented earlier. In particular, we see
that $f(q)$ exhibits a single minimum, resulting in a diode quality
factor of $\eta\approx18\%$ in the weak tilting regime with $\lambda=2$,
which is less than $\lambda_{c}\approx3.58$ {[}Fig. \ref{fig:1D-Aq-numeric}
(a, b){]}. In contrast, a strong tilt of $\lambda=4.4>\lambda_{c}$
shows two local minima in $f(q)$ and yields $\eta\approx21\%$ {[}Fig.
\ref{fig:1D-Aq-numeric} (c, d){]}. Clearly, the change in Fermiology
is correlated with a substantial enhancement of the SDE. The quantitative
values are different because we set $T=0.1$, which is quite
far from $T_{c}$, for numerical stability. 

To unearth the connection between Fermiology and the SDE more precisely,
we analytically calculate $A(q)$ in Eq. \ref{eq:Aq} in the weak
pairing limit, valid for $T$ near $T_{c}$. In this limit, Cooper
pairs predominantly form from electrons near the Fermi points. This
allows us to analytically perform the Matsubara summation and momentum
integral to obtain the following expression:
\begin{equation}
A(q)=-\sum_{i=1,2}\rho_{F}^{(i)}\left[\frac{T_{c}-T}{T_{c}}-\frac{7\zeta(3)}{16\pi^{2}T_{c}^{2}}\delta_{i}^{2}(q)\right],
\end{equation}
where $\delta_{i}(q)=(-1)^{i}\alpha q^{3}+(-1)^{i+1}3p_{F}^{(i)}\alpha q^{2}-(\lambda+(-1)^{i+1}+(-1)^{i+1}3(p_{F}^{(i)})^{2}\alpha)q+2\lambda p_{F}^{(i)}$,
and $\rho_{F}^{(i)}$ is the density of states at the $i$-th Fermi
point. For values of $|\lambda|<\lambda_{c}$,
the densities of states are given by: 
\begin{align}
\rho_{F}^{(1)} & =\left[2\pi\left(3\alpha[p_{F}^{(1)}]^{2}+(1-\lambda)\right)\right]^{-1},\nonumber \\
\rho_{F}^{(2)} & =\left[2\pi\left(3\alpha[p_{F}^{(2)}]^{2}+(1+\lambda)\right)\right]^{-1},
\end{align}
where Fermi momentum $p_{F}^{(1,2)}$ are 
\begin{align}
p_{F}^{(1)} & =\left[\frac{\mu}{2\alpha}+\sqrt{\frac{\mu^{2}}{4\alpha^2}+\frac{(1-\lambda)^3}{27\alpha^3}}\right]^{1/3}\nonumber \\
 & \quad+\left[\frac{\mu}{2\alpha}-\sqrt{\frac{\mu^{2}}{4\alpha^2}+\frac{(1-\lambda)^3}{27\alpha^3}}\right]^{1/3},\nonumber \\
p_{F}^{(2)} & =\left[-\frac{\mu}{2\alpha}+\sqrt{\frac{\mu^{2}}{4\alpha^2}+\frac{(1+\lambda)^3}{27\alpha^3}}\right]^{1/3}\nonumber \\
 & \quad+\left[-\frac{\mu}{2\alpha}-\sqrt{\frac{\mu^{2}}{4\alpha^2}+\frac{(1+\lambda)^3}{27\alpha^3}}\right]^{1/3}.
\end{align}
If $p_{F}^{(1)}+p_{F}^{(2)}\neq0$, electrons at two Fermi points
can form Cooper pairs with a finite momentum $q_{*}\approx p_{F}^{(1)}+p_{F}^{(2)}$,
where the supercurrent $j(q_{*})=0$. However, for $|\lambda|>\lambda_{c}$,
there exist three possible Fermi momenta near $p_{F,j=1,2,3}^{(2)}$,
each corresponding to a density of states $\rho_{F,j=1,2,3}^{(2)}$
for spin-up states. As illustrated in Fig. \ref{fig:1D-analytic}(d),
this leads to three potential pairing channels with electrons having
Fermi momentum near $p_{F}^{(1)}$ and spin-down, which leads to additional
structure in the free energy density. 

\begin{figure}[h]
\includegraphics[width=1.\columnwidth]{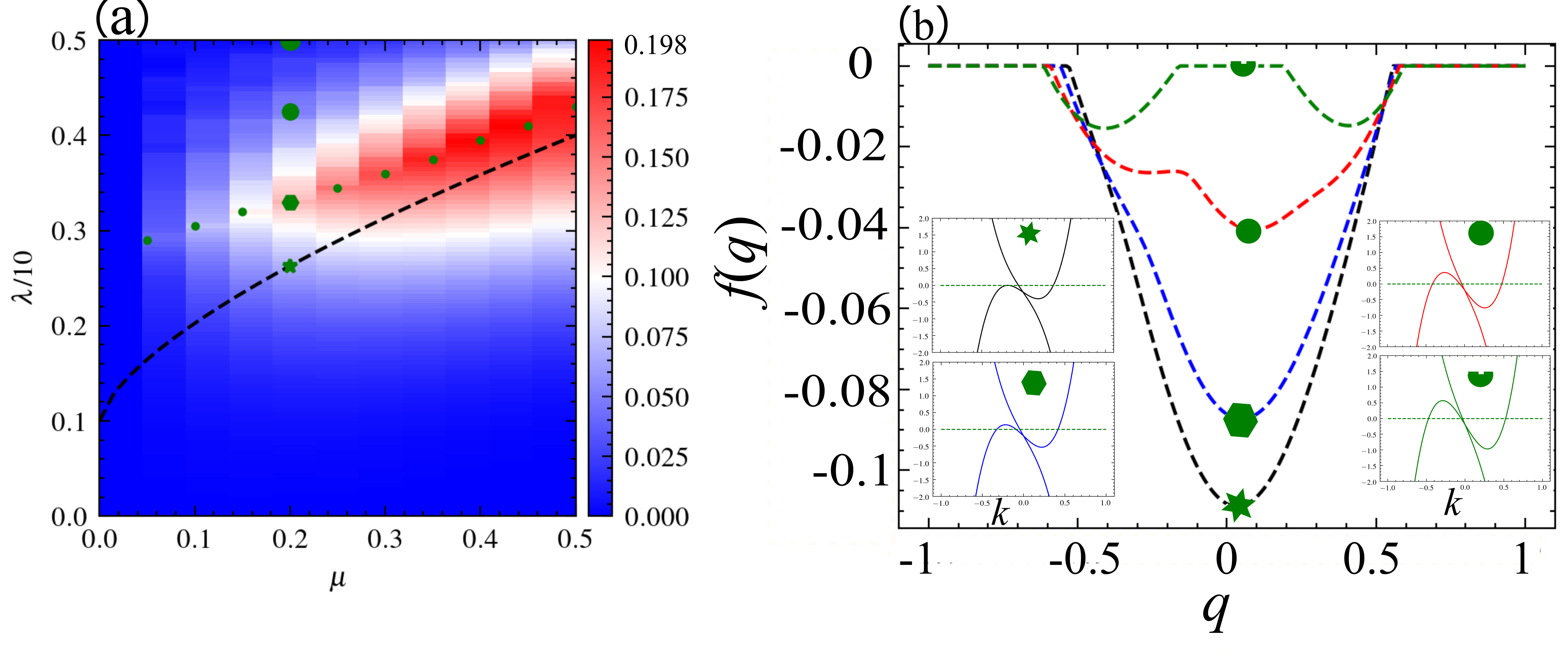} \caption{(a) The quality factor $\eta (\mu,\lambda)$ for the tilted 1D model in the $\lambda-\mu$ plane. The dashed line represents the critical tilting value $\lambda_{c}$ as a function of the chemical potential $\mu$, where $\lambda_{c}=\left|1+3\left(\frac{\mu^{2}|\alpha|}{4}\right)^{1/3}\right|$ with $\alpha=16$. The green points depict the maximum quality factor calculated numerically. (b) The free energy density with parameters corresponding to the star, hexagon, disk, and half-disk in (a). Insets display the associated band dispersion.}
\label{fig: phase} 
\end{figure}

In general, the quality factor of the SDE depends on the model's parameters. In our 1D model, two relevant parameters are $\lambda$ and $\mu$. To elucidate the relationship between the quality factor and $\left(\mu,\lambda\right)$, we present the phase diagram shown in Fig. \ref{fig: phase}(a). Interestingly, higher quality factors are observed just above the because the free energy density becomes more asymmetric near the critical line [see Fig. \ref{fig: phase}(b)]. 

We also observe that the quality factor tends to zero as $\lambda$ increases. Qualitatively, for very large $\lambda$, two Fermi points that form channel 3 in Fig. \ref{fig:1D-analytic}(d) merge into a single Fermi point [see the inset band dispersions in Fig. \ref{fig: phase}(b)]. Effectively, there are only two possible Cooper pairing channels; therefore, the diode quality factor could be diminished.

Quantitatively, we selected four typical parameters in the parameter space (denoted by star, hexagon, disk, half-disk), as shown in Fig. \ref{fig: phase}(a, b).
 At larger values of $\lambda$, the free energy density exhibits two valleys, and the two valleys are approximately mirror images of each other about the axis at $q \approx 0$. The supercurrent is defined as the derivative of the free energy density with respect to the Cooper pairing momentum. Therefore, for any positive current, there exists a negative current with the same absolute value. In other words, the diode quality factor equals zero.

Our findings not only confirm the presence of SDEs in our 1D model
with asymmetric band dispersions but also underscore the significance
of accounting for multiple Cooper pairing channels under strong tilting
conditions. The observed complex patterns in the free energy density
and supercurrent open up new avenues for optimizing superconducting
systems for non-reciprocal effects.

\section{SDE in tilted Weyl semimetals\label{sec:Weyl-semimetal}}

Weyl semimetals are intriguing materials characterized by non-degenerate
touching points, known as Weyl nodes, between their valence and conduction
bands. Weyl nodes exhibit linear dispersion and give rise to various
intriguing properties associated with the topological nature of the
bulk band structure \citep{hosur2013recent,yan2017topological,armitage2018weyl}.
There are two different types of Weyl semimetals: type I Weyl semimetals
with point-like Fermi surfaces and type II Weyl semimetals with defined
by electron and hole pockets touching at the Weyl nodes \citep{zyuzin2016intrinsic,tchoumakov2016magnetic,soluyanov2015type}.
The latter type can be obtained from the former by strongly titing
the Weyl dispersion.

In general, to realize SDEs, both $\mathcal{T}$- and $\mathcal{I}$- symmetries must be broken. The low density of states in Weyl semimetals
makes breaking the $\mathcal{T}$- and $\mathcal{I}$- symmetries easier.
On the other hand, as shown in the last section, we found that asymmetric
band dispersions can induce the SDEs. Therefore, tilted Weyl semimetals
provide us with a typical example for investigating the possibility
of realizing SDEs.

\begin{figure}[h]
\includegraphics[width=0.95\columnwidth]{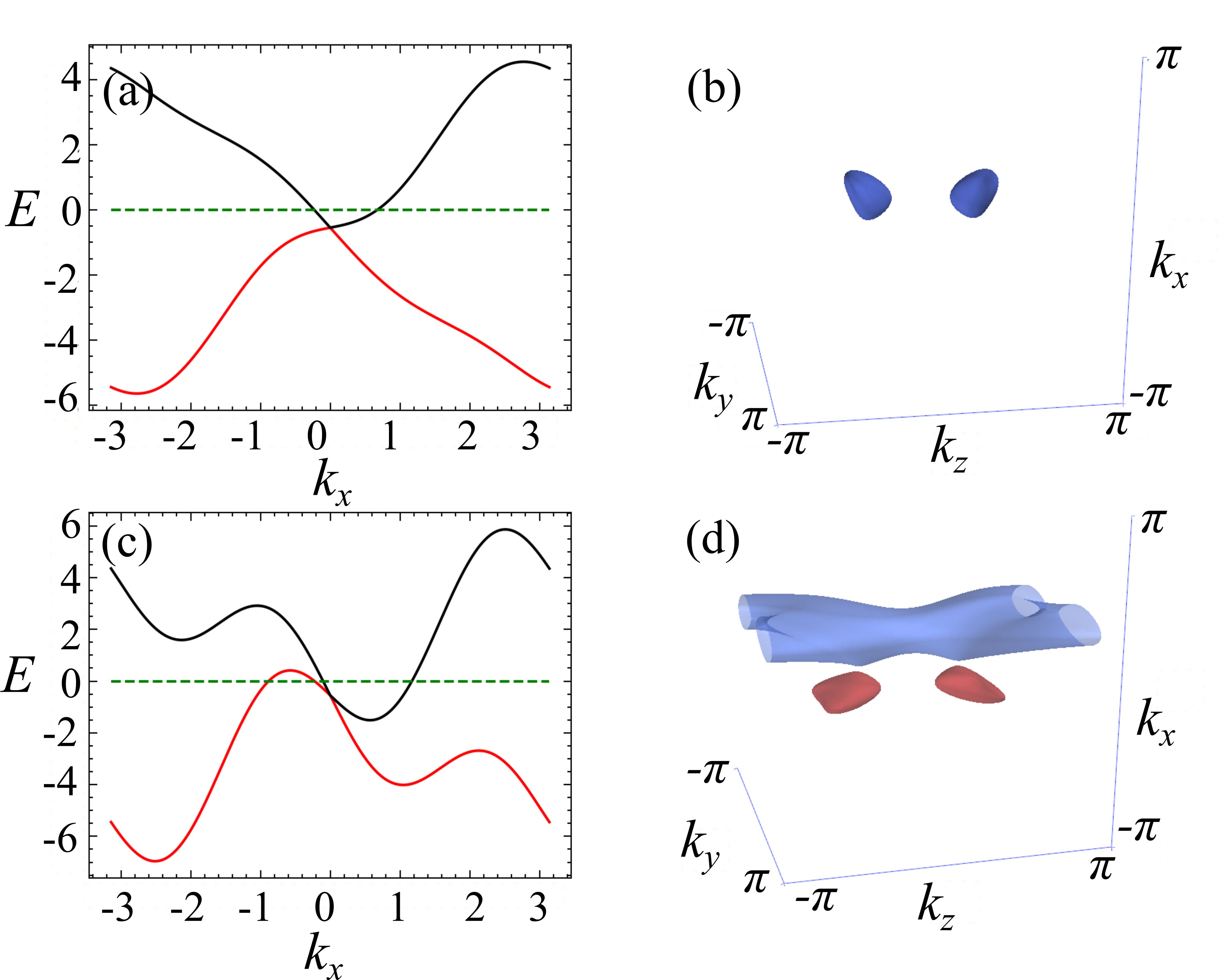} \caption{(a) Projected band structure of a tilted Weyl semimetal near the Weyl node with Bloch momentum $\mathbf{k}=(0,0,\pi/3)$. (b) Fermi surface of a weakly tilted Weyl semimetal. Parameters: $\lambda=-1/2$, $T=0$, $g=12$, and $\mu=0.55$. (c) and (d) show the same quantities as (a) and (b), respectively, with the parameter $\lambda=-2$. The band dispersion remains consistent at the other Weyl node with $\mathbf{k}=(0,0,-\pi/3).$}
\label{fig:Weyl_a} 
\end{figure}

In this section, we introduce two simple lattice models of tilted Weyl semimetals to investigate the SDEs. The Bloch Hamiltonian describing the first tilted Weyl semimetal and its corresponding energy spectrum can be expressed as follows:

\begin{align}
H_{\text{W}}(\mathbf{k}) & =\left(3+2\cos k_{z}-2\cos k_{x}-2\cos k_{y}\right)\sigma_{z}\nonumber \\
 & +2\sin  k_{+}\sigma_{x}+2\sin k_{-}\sigma_{y}+(\lambda\sin2k_{x}-\mu)\sigma_{0}\\
E_{\text{W}}^{\pm}(\mathbf{k}) & =\pm\left[\left(3+2\cos k_{z}-2\cos k_{x}-2\cos k_{y}\right)^{2}\right.\nonumber \\
 & \left.+4\sin^{2}k_{+}+4\sin^{2}k_{-}\right]^{1/2}+\lambda\sin2k_{x}-\mu
\end{align}

where the parameter $\lambda$ controls the tilt strength, $\mathbf{k}=(k_{x},k_{y},k_{z})$
represents the Bloch momentum, $\mu$ is the chemical potential, $k_{\pm}=\left(k_{x}\pm k_{y}\right)/2$, and
the Pauli matrices $(\sigma_{x},\sigma_{y},\sigma_{z})$ denote spin.
This model has two Weyl nodes at $\mathbf{k}=(0,0,\pm \pi/3)$. In Fig. \ref{fig:Weyl_a}(a, c), we provide
the eigen-energies as a function of $k_{x}$ at $k_{z}=\pi/3$, $k_{y}=0$
for the tilted Weyl semimetal with different tilt strengths. At $\lambda=0$,
the system Hamiltonian preserves $\mathcal{I}=\sigma_{z}$
but breaks $\mathcal{T} =i\sigma_{y}\mathbb{K}$. For
nonzero $\lambda$, $\mathcal{I}$- symmetry is also broken while $|\lambda|>\lambda_{c}\approx0.7$
renders the Weyl nodes type-II. For arbitrary $\lambda$ but $\mu=0$,
$H_{\text{W}}(\mathbf{k})$ obeys $\sigma_{x}H_{\text{W}}^{*}(-\mathbf{k})\sigma_{x}=-H_{\text{W}}(\mathbf{k})$,
which is particle-hole symmetry of the form (\ref{eq:chiral-anti}).
Thus, $\mu\neq0$ is necessary for a non-zero SDE. 

In the presence of s-wave pairing with a nonzero Cooper pair momentum,
the BdG Hamiltonian is given by: 
\begin{equation}
H_{\text{W}}^{\text{BdG}}(\mathbf{k},\mathbf{q})=\begin{pmatrix}H_{\text{W}}(\mathbf{k}+\mathbf{q}/2) & -i\Delta\sigma_{y}\\
i\Delta\sigma_{y} & -H_{\text{W}}^{*}(-\mathbf{k}+\mathbf{q}/2)
\end{pmatrix}
\end{equation}
The tilt is along $k_{z}$, allowing us to set $\mathbf{q}=(0,0,q)$.
$H_{\text{W}}^{\text{BdG}}(\mathbf{k},q)$ satisfies the particle-hole
symmetry $\tau_{x}\mathbb{K}H_{\text{W}}^{\text{BdG}}(\mathbf{k},q)\mathbb{K}\tau_{x}=-H_{\text{BdG}}(-\mathbf{k},q)$,
which ensures the existence of pairs of opposite eigenvalues $E_{\pm}(-\mathbf{k})$
and $-E_{\pm}(\mathbf{k})$.

\begin{figure}[h]
\includegraphics[width=0.95\columnwidth]{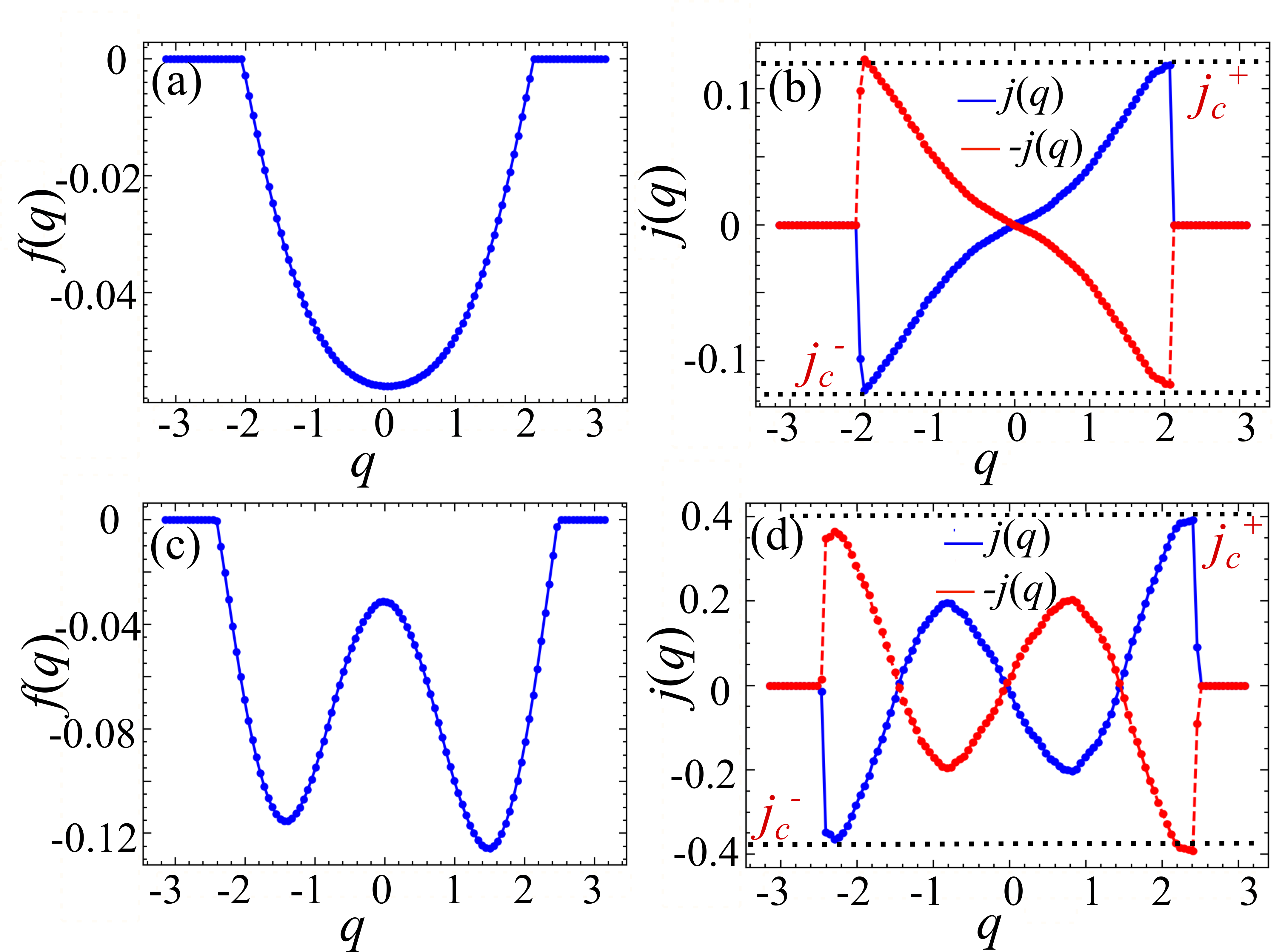} \caption{(a), (b): The free energy density $f(q)$, the supercurrent $j(q)$
(blue dotted line), and $-j(q)$ (red dotted line) for $\lambda=-1/2$,
$T=0$, $g=12$, and $\mu=0.55$. (c), (d): The same quantities as
(a, b) with the parameter $\lambda=-2$.}
\label{fig:Weyl_b} 
\end{figure}

In the 1D model, we observed that the strong tilting gives rise to
more pairing channels, which create new structures in the free energy
density and the supercurrent and enhance the SDE. In 3D model, the number
and details of the pairing channels will depend on the transverse
momenta $(k_{x},k_{y})$ in general. Nonetheless, a similar enhancement
is expected when multiple channels participate in the pairing. To
investigate this possibility, we numerically calculate $f(q)$ and
$j_{z}(q)\equiv j(q)$ at $T=0$. As shown in Fig. \ref{fig:Weyl_a}(a), for
a relatively small tilt for a given $\mu$, there is only one type
of pairing channel, only one minimum in $f(q)$ and a small difference
between $j_{c}^{\pm}$ that yield a diode quality factor of $\eta\approx1.8\%$.
However, for a larger tilted strength, three different types of Cooper
pairing channels are present, which induce two minima in $f(q)$ a
larger difference between $j_{c}^{+}$ and $j_{c}^{-}$ are boosted
diode quality factor of $\eta\approx3.7\%$ {[}see Fig. \ref{fig:Weyl_b}(c-d){]}.

We perform a similar analysis on a different lattice model of a tilted Weyl semimetal. In addition to the pockets near the Weyl nodes for the chosen parameters, there are Fermi pockets near the Brillouin zone boundary. Therefore, this model could support more possible cooper pairing channels, and the SDE  could be enhanced.

The Bloch Hamiltonian describing the 
tilted Weyl semimetal and its corresponding energy spectrum can be
expressed as: 
\begin{align}
\tilde{H}_{\text{W}}(\mathbf{k}) & =2\left(\cos k_{x}-\cos k_{0}-\cos k_{y}-\cos k_{z}+2\right)\sigma_{x}\nonumber \\
 & +2\sin k_{y}\sigma_{y}+2\sin k_{z}\sigma_{z}+(\lambda\sin2k_{z}-\mu)\sigma_{0}\\
\tilde{E}_{\text{W}}^{\pm}(\mathbf{k}) & =\pm2\left[\left(\cos k_{x}-\cos k_{0}-\cos k_{y}-\cos k_{z}+2\right)^{2}\right.\nonumber \\
 & \left.+\sin^{2}k_{y}+\sin^{2}k_{z}\right]^{1/2}+\lambda\sin2k_{z}-\mu
\end{align}

\begin{figure}[h]
\includegraphics[width=0.95\columnwidth]{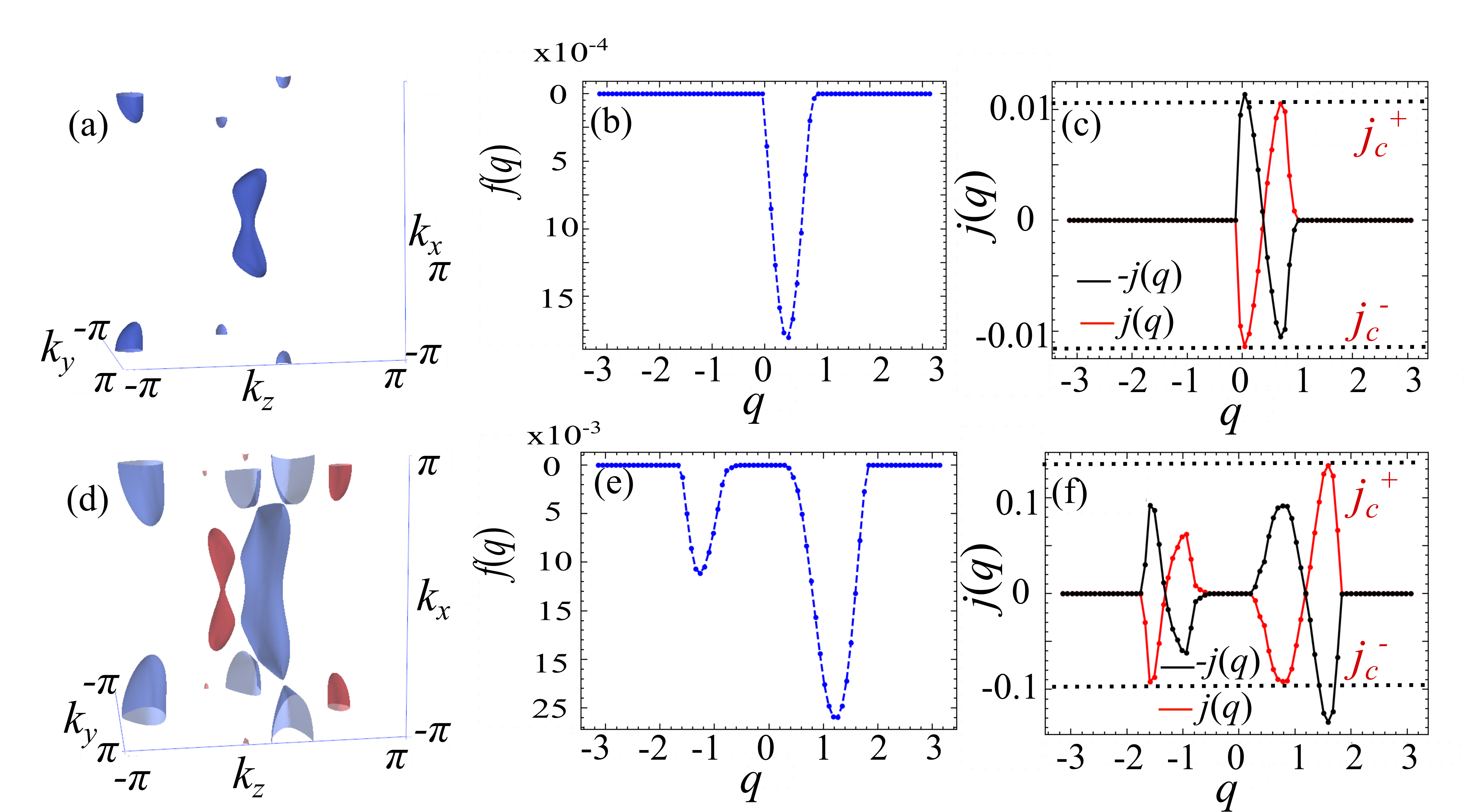} \caption{(a), (d): Fermi pockets of the tilted Weyl semimetal.
(b), (c): The free energy density $f(q)$, the supercurrent $j(q)$
(red dotted line), and $-j(q)$ (black dotted line) for $\lambda=-1$,
$T=0$, $g=10$, and $\mu=0.4$. (e), (f): The same quantities as
(b, c) with the parameter $\lambda=-2$.}
\label{fig:Weyl_c} 
\end{figure}

This model has two Weyl nodes at $\mathbf{k}=(\pm k_{0},0,0)$; we
set $k_{0}=\pi/4$ henceforth. In Fig. \ref{fig:Weyl_c}(a, d), we show the Fermi pockets
for the tilted Weyl semimetal with different tilt strengths. At $\lambda=0$,
the system Hamiltonian preserves $\mathcal{I}=\sigma_{x}$
but breaks $\mathcal{T}$- symmetry. For
nonzero $\lambda$, $\mathcal{I}$ is also broken while $|\lambda|>1$
renders the type-II Weyl nodes. For arbitrary $\lambda$ but $\mu=0$,
$\tilde{H}_{\text{W}}(\mathbf{k})$ obeys $\sigma_{z}\tilde{H}_{\text{W}}^{*}(-\mathbf{k})\sigma_{z}=-\tilde{H}_{\text{W}}(\mathbf{k})$,
which is particle-hole symmetry of the form Eq. (\ref{eq:chiral-anti}).

In the presence of s-wave pairing with a nonzero Cooper pair momentum,
the BdG Hamiltonian is given by: 
\begin{equation}
\tilde{H}_{\text{W}}^{\text{BdG}}(\mathbf{k},\mathbf{q})=\begin{pmatrix}\tilde{H}_{\text{W}}(\mathbf{k}+\mathbf{q}/2) & -i\Delta\sigma_{y}\\
i\Delta\sigma_{y} & -\tilde{H}_{\text{W}}^{*}(-\mathbf{k}+\mathbf{q}/2)
\end{pmatrix}
\end{equation}

The tilt is along $k_{z}$, allowing us to set $\mathbf{q}=(0,0,q)$.
$\tilde{H}_{\text{W}}^{\text{BdG}}(\mathbf{k},q)$ satisfies the particle-hole
symmetry $\tau_{x}\mathbb{K}\tilde{H}_{\text{W}}^{\text{BdG}}(\mathbf{k},q)\mathbb{K}\tau_{x}=-\tilde{H}_{\text{BdG}}(-\mathbf{k},q)$,
which ensures the existence of pairs of opposite eigenvalues $\tilde{E}_{\pm}(-\mathbf{k})$
and $-\tilde{E}_{\pm}(\mathbf{k})$.

As shown in Fig. \ref{fig:Weyl_c}(b-c), for
a relatively small tilt for a given $\mu$, there is only one minimum in $f(q)$ and a small difference
between $j_{c}^{\pm}$ that yield a diode quality factor of $\eta\approx3.8\%$.
However, for a larger tilted strength, two minima in $f(q)$ a
larger difference between $j_{c}^{+}$ and $j_{c}^{-}$ are boosted
diode quality factor of $\eta\approx18.4\%$ {[}see Fig. \ref{fig:Weyl_c}(e-f){]}. 
The quality factor of the SDE in this model is much higher than the diode quality factor in the first model, confirming that multiple Cooper pairing channels can enhance the diode quality factor.

\section{SDE in tilted Dirac semimetals\label{sec:Dirac-semimetal}}

Similar to Weyl semimetals, in a Dirac semimetal, the valence and
conduction bands touch linearly at specific points in the Brillouin
zone, known as Dirac points, where the energy dispersion relation
is linear in momentum \citep{young2012dirac,young2015dirac,gibson2015three}.
The existence of these three-dimensional Dirac points is of profound
significance in condensed matter physics. At the quantum critical
point, where a transition occurs between a normal insulator and a
topological insulator, a three-dimensional Dirac semimetal manifests 
\citep{murakami2007phase}. This quantum critical point represents
a delicate balance between different electronic states, resulting
in the appearance of a Dirac semimetal phase that possesses distinct
topological properties. The formation of this exotic phase further
highlights the role of symmetries in dictating the behavior of electronic
states and their topological nature.

In the last section, we have shown that SDE could be realized in tilted
Weyl semimetals. Due to the similarity between Weyl semimetals and
Dirac semimetals, a natural question arises: can introducing a perturbation
term to the Dirac semimetal, which tilts the band dispersion and breaks
both $\mathcal{T}$- and $\mathcal{I}$- symmetries, support the emergence
of SDEs? To answer this question, we consider a lattice model of the
Dirac semimetals and study the possibility of SDEs induced by the
tilting. 

\begin{figure}[h]
\includegraphics[width=0.95\columnwidth]{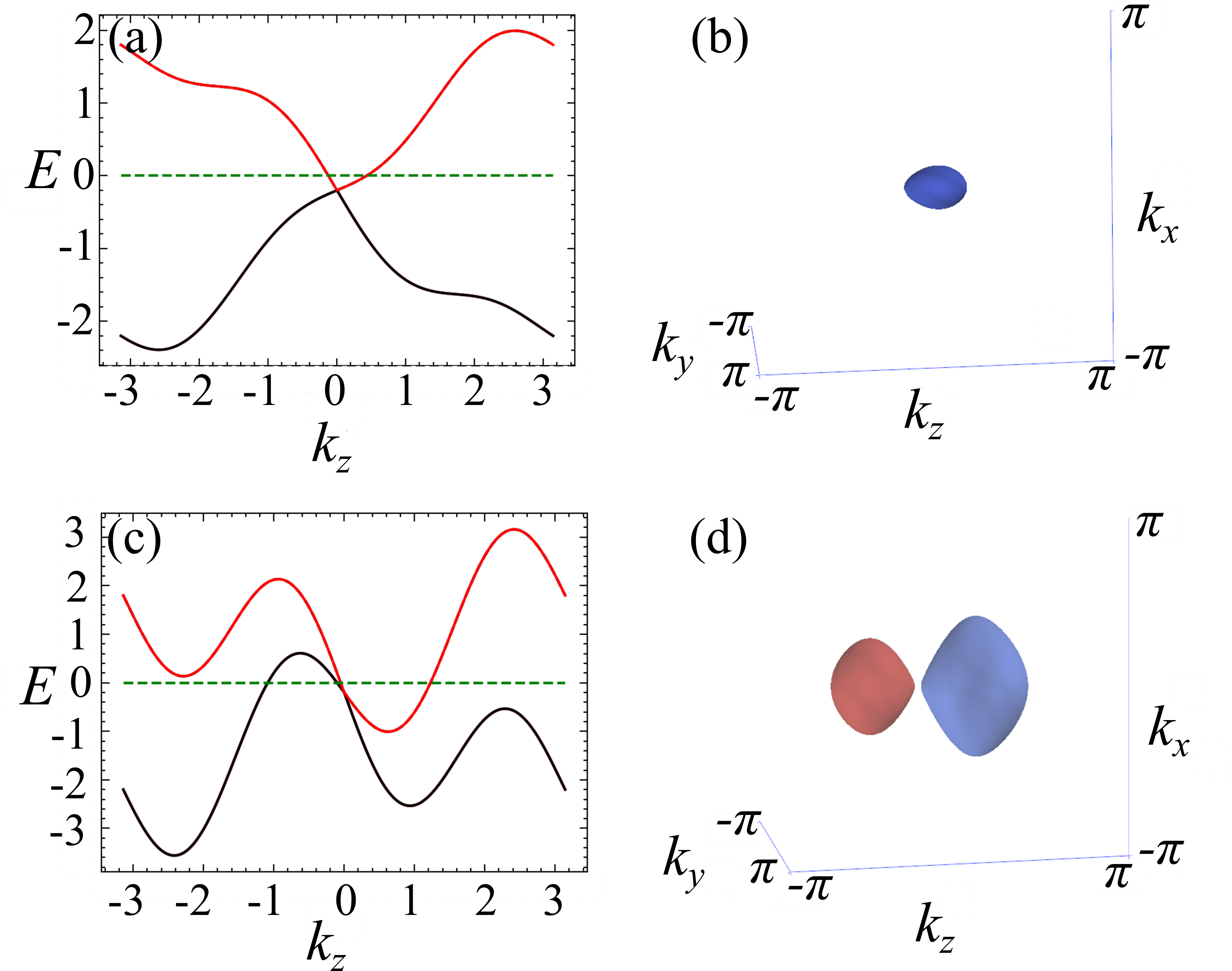} \caption{(a) Projected band structure of the Dirac semimetal. (b) Fermi surface for $\lambda=-0.3$, $T=0$, $g=2.6$, and $\mu=0.2$. (c), (d) Same quantities as in (a, b), with the parameters being identical to those in (a, b), except for the parameter $\lambda=-1.5$.}
\label{fig:Dirac_a} 
\end{figure}

We focus on a cubic lattice model with a single Dirac point at the $\Gamma=(0,0,0)$ point. The dispersion is tilted in a specific direction, assumed to be in the $z$ direction as shown in Fig. \ref{fig:Dirac_a}. The Bloch Hamiltonian is:

\begin{align}
H_{\text{D}}(\mathbf{k}) & =\sin k_{x}\Gamma_{zy}+\sin k_{y}\Gamma_{zx}+\sin k_{z}\Gamma_{y0}\nonumber \\
 & +(3-\cos k_{x}-\cos k_{y}-\cos k_{z})\Gamma_{x0}\nonumber \\
 & +(\lambda\sin k_{z}-\mu)\Gamma_{00}
\end{align}
where the matrix $\Gamma_{ab}\equiv\tau_{a}\otimes\sigma_{b}$ with
$a$, $b$ $\in(0,x,y,z)$. The term proportional to $\lambda$ induces
tilting and breaks the $\mathcal{T}$- and $\mathcal{I}$- symmetries
while a non-zero $\mu$ is needed to break symmetries studied in Sec.
\ref{sec:symmetry}.

\begin{figure}[h]
\includegraphics[width=0.95\columnwidth]{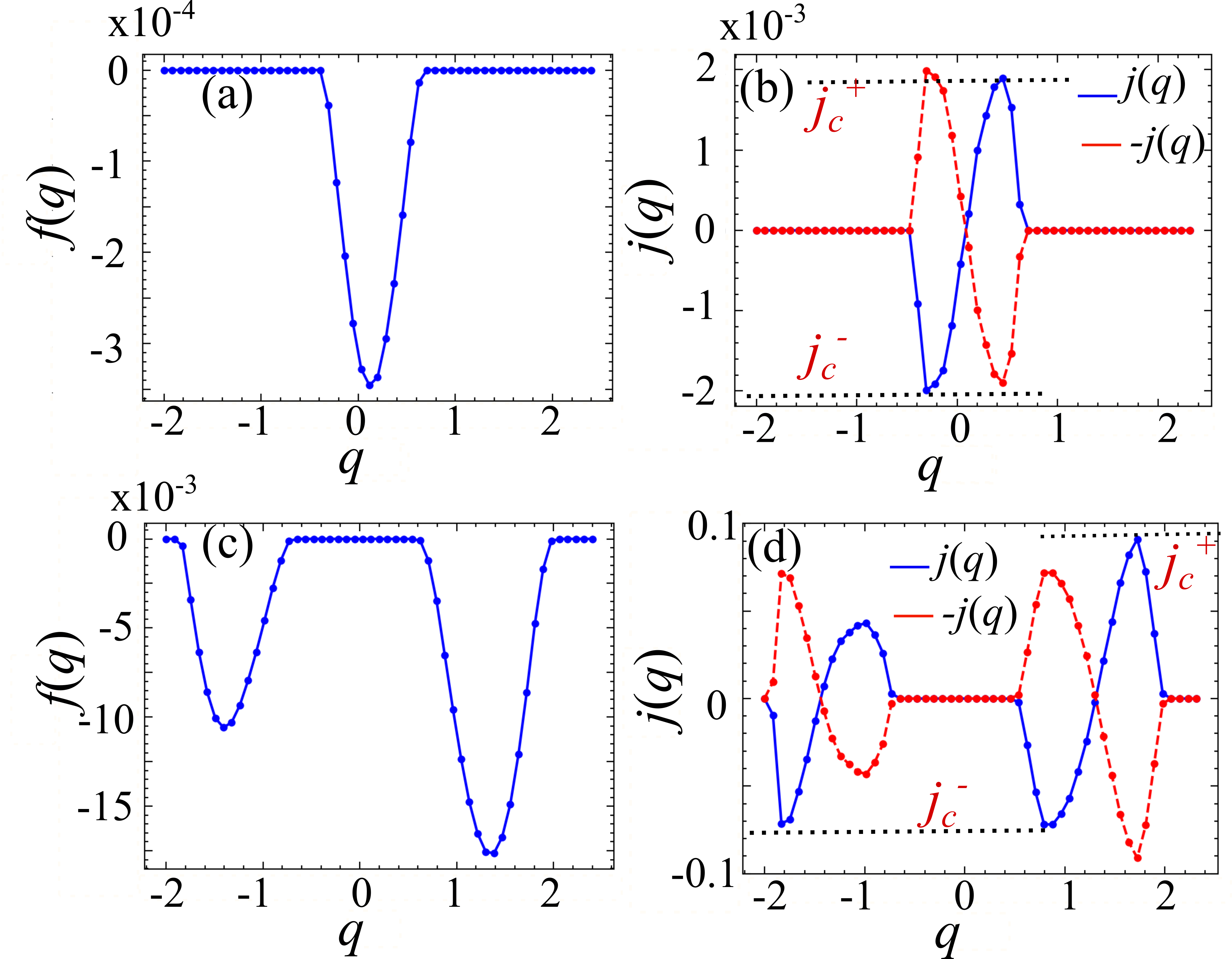} \caption{(a), (b): Free energy density $f(q)$, supercurrent $j(q)$ (blue dotted line), and $-j(q)$ (red dotted line) for parameters $\lambda=-0.3$, $T=0$, $g=2.6$, and $\mu=0.2$. (c), (d): Same quantities as in (a, b) with identical parameters, except for $\lambda=-1.5$.}
\label{fig:Dirac_b} 
\end{figure}

$s$-wave superconductivity is captured by the BdG Hamiltonian: 
\begin{equation}
H_{\text{D}}^{\text{BdG}}(\mathbf{k},\mathbf{q})=\begin{pmatrix}H_{\text{D}}(\mathbf{k}+\mathbf{q}/2) & -i\Delta\sigma_{y}\\
i\Delta\sigma_{y} & -H_{\text{D}}^{*}(-\mathbf{k}+\mathbf{q}/2)
\end{pmatrix}
\end{equation}
As demonstrated in Fig. \ref{fig:Dirac_b}, our investigation reveals
intriguing similarities between the free energy density and the SDEs
observed in Dirac semimetals and those previously observed in Weyl
semimetals. The quality factor $\eta\approx2.5\%$ at weak tilting
with $\lambda=-0.3$ and $\eta\approx11.7\%$ at stronger tilting
with $\lambda=-1.5$. This enhancement is accompanied by the appearance
of multiple pairing channels and multiple minima in the free energy.
These behaviors motivate exploring tilted Dirac semimetals as well
for the realization of SDEs. 

\section{Candidate materials}

For materials with broken $\mathcal{T}$- and $\mathcal{I}$- symmetries,
the realization of SDEs might be hindered by additional lattice symmetries,
such as mirror symmetry or reflection symmetry. Consequently, these
additional symmetries would also need to be broken to enable the occurrence
of SDEs. One such material exemplifying this is Ti$_{2}$MnAl, with
space group $F\bar{4}3M$ (No. 216) \citep{PhysRevB.97.060406}. In
Ti$_{2}$MnAl, weak spin-orbit coupling further breaks the mirror
symmetry (M$_{\pm110}$), leading to different tilts between the two
mirror-symmetric Weyl points. Another set of materials can be found
in the RAlX family with the space group I4$_{1}$md (No. 109), where
R represents rare earth metals like Pr, Ce, and Sm, and X denotes
Ge or Si \citep{sanchez2020observation,chang2018magnetic}. These
materials lack horizontal mirror symmetry, which increases the likelihood
of asymmetric bands in the z-direction. If superconductivity could
be realized in them, then they are potential candidate materials for
verifying our theoretical studies. 

\section{CONCLUSIONS}

In this work, we delved into the intriguing phenomenon of SDEs in topological semimetals.
We demonstrated, by investigating a simple 1D toy model using various numerical and analytical methods, that multiple pairing channels rather than tilting the dispersion enrich the superconducting physics and enhance the SDE. We carried this understanding to 3D Weyl and Dirac semimetals, showed the existence of the SDE in these systems, and demonstrated its enhancement due to multiple Fermi pockets and pairing channels.

Our findings hold implications for future explorations of superconducting
phenomena and topological effects in condensed matter systems. Moreover,
the intrinsic nature of SDEs in the presence of asymmetric band
dispersions suggests a promising avenue for designing advanced superconducting
devices and harnessing nonreciprocal transport in quantum technologies.
Ultimately, this research opens up new directions for investigating
emergent phenomena at the intersection of superconductivity and topological
physics.

\acknowledgments

This work was supported by the Department of Energy grant no. DE-SC0022264.

\bibliographystyle{apsrev4-1}
\bibliography{sdlib_1}

\end{document}